\documentclass{aastex}
\usepackage{spr-astr-addons}
\usepackage{url}
\usepackage{eurosym}
\usepackage{amssymb}
\usepackage{amsfonts}
\usepackage{amsmath}
\usepackage{graphicx}
\RequirePackage{color}

\newcommand{\emaila}{}

\begin{document}

\title{Gravitinos Tunneling From Traversable Lorentzian Wormholes}

\shorttitle{Gravitinos Tunneling From TLWH}
\shortauthors{Sakalli et al.}

\author{I. Sakalli \altaffilmark{1} \emaila{izzet.sakalli@emu.edu.tr }}  \and \author{A. Ovgun  \altaffilmark{1}\emaila{ali.ovgun@emu.edu.tr } }
\affil{Physics Department , Eastern Mediterranean University, Famagusta, Northern
Cyprus,  Mersin 10, Turkey.}

\date{\today}
\begin{abstract}
Recent research shows that Hawking radiation (HR) is also possible around
the trapping horizon of a wormhole. In this article, we show that the HR of
gravitino (spin-$3/2$) particles from the traversable Lorentzian wormholes
(TLWH) reveals a negative Hawking temperature (HT). We first introduce the TLWH in the past outer trapping
horizon geometry (POTHG). Next, we derive the Rarita-Schwinger equations (RSEs) for that
geometry. Then, using both the Hamilton-Jacobi (HJ) ans\"{a}tz and the WKB
approximation in the quantum tunneling method, we obtain the probabilities
of the emission/absorption modes. Finally, we derive the tunneling rate of
the emitted gravitino particles, and succeed to read the HT of the TLWH.
\end{abstract}

\section{Introduction}

An interesting phenomenon that corresponds to spontaneous emissions (as if a
black body radiation) from a black hole (BH) is the HR. It is a semi-classical outcome of the quantum field theory \citep{Haw,Haw1}. HR dramatically changed our way of looking
to the BHs; they are not absolutely black and cold objects, rather they emit
energy with a characteristic temperature: HT. Event horizon,
where is an irreversible point (in classical manner) for any object including
photons is the test-bed of the gedanken experiment for the HR. The studies
concerning this phenomenon have been carrying on by using different methods.
In particular, the quantum tunneling \citep{PW} of particles with different
spins from the various BHs have gained momentum in the recent years (the
reader may be referred to
\citep{Vaz,Jing,kerner1,kerner2,kerner3,Mann,yang,sharifa,Kruglova,ran2,sharif,ran,ChenZhou,ali2,ali1,ali3,sucu,gohar1,huan,singh,MDehghani,hale}
and references cited therein). Recently, it has been shown that HR of the
bosons with spin-0 (scalar particles) and spin-1 (vector particles) from the
TLWH  \citep{wh}, which is a bridge or tunnel between different regions of the spacetime
is possible by using the POTHG
\citep{Diaz,Diaz1,aliwh}. Wormhole has been extensively studied in different areas \citep{garattini,kuhfittig1,rahaman1,rahaman2,rahaman3,halilsoy}. However, HT of the TLWH appears to be negative
because of the phantom energy (exotic matter: the sum of the
pressure and energy density is negative) that supports the broadness of the
wormhole throat \citep{wh}. In addition, it is a well-known fact that the virtual particle-antiparticle pairs are created near the horizon. In a BH spacetime the real particles with positive energy and temperature are emitted towards spatial infinity \citep{wald}. However, in the POTHG which is analog to the white hole geometry, the antiparticles come out from the horizon \citep{Helou2}. In other words, our analysis predicts that the energy spectrum of the antiparticles leads to a negative temperature for the TLWH. For the subject of the white hole radiation, the reader may refer to \citep{peltola}.  

As it is shown by Caldwell et al \citep{Cald}, the dark
matter (DM) \citep{dm} could have a phantom energy. In this regard, the phantom energy
can keep apart every bound object until the Cosmos eventuates in the Big-Rip \citep{BRP}. On
the other hand, DM does not emit, reflect or absorb light, making it not just
dark but entirely transparent. But if the DM particles strolling around a BH or
\ a wormhole can produce gamma-rays would give a possibility to study the radiation of this
mysterious matter \citep{gamma1,dm1}. DM has many candidates, and
gravitino (spin-$3/2$) is one of them \citep{Kawas,David}. Gamma-ray decay of
the gravitino DM has been very recently studied in \citep{dm1}. So, HR
of the gravitinos from the BHs and/or wormholes could make an impact on the
production of the DM. Behaviors of the gravitino's wave function are governed
by the RSEs \citep{Mann,rsch1}. So, our main motivation in this paper is to
investigate the HR of the gravitino tunneling from the TLWH geometry. Using
the RSEs and HJ method, we aim to regain the standard HT\ of the TLWH.

The structure of this paper is as follows. In Sec. II, we introduce the 3+1
dimensional TLWH \citep{Diaz1} and analyzes the RSEs for the gravitino
particles in the POTHG of the TLWH
\citep{hawyard,hayward1,hayward2,MisnerSharp,Topology}. We show that the RSEs
are separable when a suitable HJ \"{a}nsatz is employed. Then the radial equation can
be reduced to a coefficient matrix equation that makes us possible to compute
the probabilities of the emission/absorption of the gravitinos. Finally, we
calculate the tunneling rate of the radiated gravitinos, and retrieve the standard HT
of the TLWH. We summarize and discuss our results in Sec. III.

\section{Quantum Tunneling of Gravitinos From 3+1 Dimensional TLWH}

For the wave equation of the gravitino (spin-3/2) particles, we start with
the massless (the mass has no remarkable effect in the computation of the
quantum tunneling \citep{Mann}) RSEs \citep{rsch1,ferm,ferm1,ferm2}:
\begin{gather}
i\gamma^{\nu}\left(  D_{\nu}\right)  \Psi_{\mu}=0,\label{1}\\
\gamma^{\mu}\Psi_{\mu}=0,\label{2nn}%
\end{gather}

where $\Psi_{\mu}\equiv\Psi_{\mu a}$ is a vector-valued spinor and the
$\gamma^{\mu}$ matrices satisfy $\left\{  \gamma^{\mu},\gamma^{\nu}\right\}
=2g^{\mu\nu}$.\ The first equation is the Dirac equation applied to every
vector index of $\Psi$, while the second is a set of additional constraints to
ensure that no ghost state propagates; that is, to ensure that $\Psi$
represents only spin-$3/2$ fermions, with no spin-$1/2$ mixed states
\citep{Mann,ferm}.

The covariant derivative obeys%

\begin{equation}
D_{\mu}=\partial_{\mu}+\frac{i}{2}\Gamma_{\mu}^{\alpha\beta}J_{\alpha\beta},
\label{3}%
\end{equation}

where%

\begin{align}
\Gamma_{\mu}^{\alpha\beta}  &  =g^{\beta\gamma}\Gamma_{\mu\gamma}^{\alpha
},\nonumber\\
J_{\alpha\beta}  &  =\frac{i}{4}\left[  \gamma^{\alpha},\gamma^{\beta}\right]
,\nonumber\\
\{\gamma^{\alpha},\gamma^{\beta}\}  &  =2g^{\alpha\beta}\times I. \label{4}%
\end{align}

The metric of TLWH in the generalized retarded Eddington-Finkelstein
coordinates (REFCs), which is the POTHG, is given by \citep{Diaz1}%

\begin{equation}
ds^{2}=-Fdu^{2}-2dudr+r^{2}\left(  d\theta^{2}+\sin^{2}\theta d\varphi
^{2}\right)  , \label{5}%
\end{equation}

where $F=$ $1-2M/r$. Misner-Sharp energy is represented by $M=\frac{1}%
{2}r(1-\partial^{a}r\partial_{a}r)$\ which becomes $M=\frac{1}{2}r_{h}$ on the
trapping horizon ($r_{h}$) \citep{MisnerSharp}. Marginal surfaces having
$F(r_{h})=0$ are the past marginal surfaces in the REFCs \citep{Diaz}.

For solving the RSEs, we use the following Dirac $\gamma$-matrices:%
\begin{align}
\gamma^{u}  &  =\frac{1}{\sqrt{F}}\left(
\begin{array}
[c]{cc}%
-i & -\sigma^{3}\\
-\sigma^{3} & i
\end{array}
\right)  ,\,\,\ \gamma^{r}=\sqrt{F}\left(
\begin{array}
[c]{cc}%
0 & \sigma^{3}\\
\sigma^{3} & 0
\end{array}
\right)  ,\nonumber\\
\gamma^{\theta}  &  =\frac{1}{r}\left(
\begin{array}
[c]{cc}%
0 & \sigma^{1}\\
\sigma^{1} & 0
\end{array}
\right)  ,\,\,\,\ \gamma^{\phi}=\frac{1}{r\text{sin}\theta}\left(
\begin{array}
[c]{cc}%
0 & \sigma^{2}\\
\sigma^{2} & 0
\end{array}
\right)  , \label{6}%
\end{align}

where the Pauli matrices are given by%

\begin{equation}
\sigma^{1}=\left(
\begin{array}
[c]{cc}%
0 & 1\\
1 & 0
\end{array}
\right)  ,\sigma^{2}=\left(
\begin{array}
[c]{cc}%
0 & -i\\
i & 0
\end{array}
\right)  ,\sigma^{3}=\left(
\begin{array}
[c]{cc}%
1 & 0\\
0 & -1
\end{array}
\right)  . \label{7}%
\end{equation}

Gravitino wave function ($\psi$) has two spin states [spin up (i.e. positive
$r-$direction) and spin down (i.e. negative $r-$direction)]:%

\begin{equation}
\psi_{\nu\uparrow}=\left(  a_{\nu},0,c_{\nu},0\right)  e^{\frac{i}{\hbar
}S_{\uparrow}(u,r,\theta,\phi)}, \label{8}%
\end{equation}

\begin{equation}
\psi_{\nu\downarrow}=\left(  0,b_{\nu},0,d_{\nu}\right)  e^{\frac{i}{\hbar
}S\downarrow(u,r,\theta,\phi)}, \label{9}%
\end{equation}

where $S(u,r,\theta,\phi)$ denotes the gravitino action which is going to be
expanded in powers of $\hbar$, and $ a_{\nu},b_{\nu},c_{\nu},d_{\nu
} $ are the arbitrary constants. Here we shall only consider the spin
up case, since the spin down case is fully analogous with it. The action for
the spin-up states can be chosen as follows
\begin{align}
S_{\uparrow}(u,r,\theta,\phi)  &  =S_{\uparrow0}(u,r,\theta,\phi)+\hbar
S_{\uparrow1}(u,r,\theta,\phi)\nonumber\\
&  +\hbar^{2}S_{\uparrow2}(u,r,\theta,\phi)+.... \label{10}%
\end{align}

Therefore, the corresponding RSEs become
\begin{align}
\frac{1}{\sqrt{F}}\left[  \left(  ia_{0}-c_{0}\right)  (\partial
_{u}S_{\uparrow0})\right]  +\sqrt{F}(-c_{0}\partial_{r}S_{\uparrow0})  &
=0,\label{11}\\
\frac{1}{r\text{sin}\theta}(-ic_{0}\partial_{\phi}S_{\uparrow0})+\frac{1}%
{r}(-c_{0}\partial_{\theta}S_{\uparrow0})  &  =0,\label{12}\\
\frac{1}{\sqrt{F}}\left[  \left(  -a_{0}-ic_{0}\right)  (\partial
_{u}S_{\uparrow0})\right]  +\sqrt{F}(-a_{0}\partial_{r}S_{\uparrow0})  &
=0,\label{13}\\
\frac{1}{r\text{sin}\theta}(-ia_{0}\partial_{\phi}S_{\uparrow0})+\frac{1}%
{r}(-a_{0}\partial_{\theta}S_{\uparrow0})  &  =0, \label{14}%
\end{align}

with the constraints equations:%

\begin{align}
\frac{-a_{0}+c_{0}}{\sqrt{F}}+\sqrt{F}c_{1}+\frac{d_{2}}{r}-\frac{id_{3}%
}{r\text{sin}\theta}  &  =0\label{15}\\
\frac{-d_{0}}{\sqrt{F}}-\sqrt{F}d_{1}+\frac{c_{2}}{r}+\frac{ic_{3}%
}{r\text{sin}\theta}  &  =0\label{16}\\
\frac{a_{0}}{\sqrt{F}}+\sqrt{F}a_{1}+\frac{b_{2}}{r}-\frac{ib_{3}}%
{r\text{sin}\theta}  &  =0\label{17}\\
\frac{-b_{0}+id_{0}}{\sqrt{F}}-\sqrt{F}b_{1}+\frac{a_{2}}{r}+\frac{ia_{3}%
}{r\text{sin}\theta}  &  =0 \label{18n}%
\end{align}
\newline

Equations (15-18) are not important here. Because these equations give an
independent wave solution, so that they have no effect on the action
\citep{Mann}.

Afterwards, the separation of variables method is applied to the action
$S_{\uparrow0}(u,r,\theta,\phi)$:
\begin{equation}
S_{\uparrow0}=Eu-W(r)-j_{\theta}\theta-j_{\phi}\phi+K, \label{19n}%
\end{equation}
where $E$ and $\left(  j_{\theta},j_{\phi}\right)  $ are energy and angular
constants, respectively. However, $K$ is an arbitrary complex constant. Thus,
Eqs. (11-14) reduce to%

\begin{gather}
-\frac{ia_{0}}{\sqrt{F}}E+\frac{c_{0}}{\sqrt{F}}E-c_{0}\sqrt{F}W^{\prime
}=0,\label{20}\\
\frac{-c_{0}}{r}\left(  j_{\theta}+\frac{i}{\text{sin}\theta}j_{\phi}\right)
=0,\label{21}\\
\frac{a_{0}}{\sqrt{F}}E+\frac{ic_{0}}{\sqrt{F}}E-a_{0}\sqrt{F}W^{\prime
}=0,\label{22}\\
\frac{-a_{0}}{r}\left(  j_{\theta}+\frac{i}{\text{sin}\theta}j_{\phi}\right)
=0. \label{23n}%
\end{gather}

Equations (21) and (23) are about the solutions of $\left(  j_{\theta}%
,j_{\phi}\right)  $, and they do not have contribution to the tunneling rate.
For this reason, we simply ignore them. Namely, the master equations for the tunneling
rate are Eqs. (20) and (22). To analyze them, we first consider the case of
$a_{0}=ic_{0}$ \citep{yian}. Using Eqs. (20) and (22), we now have a solution
for $W(r)$ as
\begin{equation}
W_{1}=\int\frac{2E}{F}dr. \label{24n}%
\end{equation}

The integrand has a simple pole at $r=r_{h}$. Choosing the contour as a half
loop going around this pole from left to right and integrating, one obtains%

\begin{equation}
W_{1}=\frac{i2\pi E}{F^{\prime}(r_{h})}=\frac{i\pi E}{\kappa|_{H}}.
\label{25n}%
\end{equation}

where $\kappa|_{H}=\left.  \partial_{r}F/2\right\vert _{r=r_{H}}$ is the
surface gravity at the horizon. On the other hand, if one sets $a_{0}=-ic_{0}%
$, this time Eqs. (20) and (22) admit the following solution for $W(r):$%

\begin{equation}
W_{2}=0. \label{26n}%
\end{equation}

Hence, we can derive the ingoing/outgoing imaginary action solutions as%

\begin{equation}
\operatorname{Im}S_{1}=\operatorname{Im}W_{1}+\operatorname{Im}K, \label{27n}%
\end{equation}

\begin{equation}
\operatorname{Im}S_{2}=\operatorname{Im}W_{2}+\operatorname{Im}%
K=\operatorname{Im}K. \label{28}%
\end{equation}

We can now set $S_{1}$ for the action of absorbed (ingoing) gravitinos. We can
tune their probability:%

\begin{equation}
\Gamma_{in}=\exp(-2\operatorname{Im}S_{1}), \label{29n}%
\end{equation}

to \%100 by letting%

\begin{equation}
K=-\frac{i\pi E}{\kappa|_{H}}. \label{30n}%
\end{equation}

Consequently, the probability of the emitted (outgoing) gravitinos becomes%

\begin{equation}
\Gamma_{out}=\exp(-2\operatorname{Im}S_{2})=\exp(\frac{2\pi E}{\kappa|_{H}}).
\label{31n}%
\end{equation}

Recalling the definition of the tunneling rate:%

\begin{equation}
\Gamma=\frac{\Gamma_{out}}{\Gamma_{in}}=\exp(\frac{2\pi E}{\kappa|_{H}}),
\label{32n}%
\end{equation}

which is also equivalent to the Boltzmann factor: $\Gamma=\exp(-E/T)$, we read
the HT of the TLWH as follows%

\begin{equation}
T_{H}=-\frac{\kappa|_{H}}{2\pi}, \label{33n}%
\end{equation}
which is a negative temperature. This result implies that if the trapping
horizon remains in the past outer region, the wormhole throat would have a
negative temperature \citep{Diaz1}. The phantom energy, which is the special
case of the exotic matter could be the reason of that negative temperature %
\citep{Diaz,Diaz1,aliwh,Diaz2,Saridakis,de1,Helou}. On the other hand, when $%
K=0$ in the action (19), it is possible to obtain the positive temperature: $%
T_{H}=+\frac{\kappa |_{H}}{2\pi }$. Although, the latter remark contradicts with the previous results \citep{Diaz1,aliwh} (and whence, one may easily get rid of the case of $K=0$), however Hong and Kim \citep{hong} showed that possibility of negative/positive temperature of the wormhole depends on the exotic matter distribution.

\section{Conclusion}

In this work, we have studied the HR of the gravitino particles from the
TLWH in 3+1 dimensions. TLWH has been introduced in the POTHG. We have
analyzed the RSEs in the background of the TLWH with the help of HJ method.
The probabilities of the emitted/absorbed gravitino particles from the
trapped horizon of the TLWH have been computed. After comparing the obtained
tunneling rate with the Boltzmann factor, we have recovered the standard HT
of the TLWH, which is a negative temperature. This is the special condition
in which the high-energy states are more occupied than lower-energy states 
\citep{sci1}. Another possibility of the negative temperature may originate
from the exotic matter distribution of the wormhole \citep{hong}. Meanwhile, very recently it has been claimed by \cite{Helou2} that HR does not occur in the POTHG. In fact, the latter debatable remark is based on the study of \cite{ellis} stating that cosmic matter flux may turn the HR off. On the other hand, Hayward show that switching off the radiation causes the wormhole to collapse to a Schwarzschild BH \citep{hayward3}. 

In summary, gravitinos can tunnel through wormhole [simply this can be thought as a wormhole with one
entrance (BH) and one exit (white hole)]. In such a case, gravitinos tunnel from the BH with positive temperature, while they
tunnel through the white hole with negative temperature. Thus our calculations are based on the exit of the wormhole, just as the white hole case.  
Besides, we have shown that positive temperature can be obtained by tuning
the $K$-constant in the action (19) to zero. However, the latter result is a
debatable issue, and it demands much deeper analysis. This will be our next
venture in this line of study.

\acknowledgments
We would like to thank the editor and the anonymous referee for their comments and suggestions.

\end{document}